\documentclass[final,3p,times,twocolumn,natbib]{elsarticle}
\usepackage{graphicx}
\usepackage{natbib}
\usepackage{hyperref}
\usepackage{ulem}
\usepackage{amssymb}
\usepackage{lineno}
\usepackage{xcolor}

\journal{Physics Letter B}

\begin{document}

\begin{frontmatter}


\title{Unraveling the reaction mechanism for large alpha production and incomplete fusion in reactions involving weakly bound stable nuclei}% in heavy-ion induced reactions

%\title{Dominance of cluster-stripping in large alpha production and incomplete fusion \\in reactions involving weakly bound $^7$Li}% in heavy-ion induced reactions


\author[a,b]{S.~K.~Pandit}
\ead{sanat@barc.gov.in}
\author[a,b]{A.~Shrivastava}
\author[a,b]{K. Mahata}
\author[c]{N. Keeley}
\author[a,b]{V.~V.~Parkar}
\author[d]{R.~Palit}
\author[a,b]{P.~C.~Rout}
\author[a]{K.~Ramachandran}

\author[a]{A.~Kumar}
\author[e]{S.~Bhattacharyya}
\author[d]{V.~Nanal}

\author[d]{S.~Biswas}
\author[d]{S.~Saha}
\author[d]{J.~Sethi}
\author[d]{P.~Singh}
\author[f]{S.~Kailas}

\address[a]{Nuclear Physics Division, Bhabha Atomic Research Centre, Mumbai - 400085, India}
\address[b]{Homi Bhabha National Institute, Anushaktinagar, Mumbai - 400094, India}
\address[c]{National~Centre~for~Nuclear~Research,~ul.~Andrzeja~So\l tana~7,~05-400~Otwock,~Poland}
\address[d]{Department of Nuclear and Atomic Physics, Tata Institute of Fundamental Research, Mumbai 400005, India}
\address[e]{Variable Energy Cyclotron Centre, Kolkata - 700064, India}
\address[f]{UM-DAE Centre for Excellence in Basic Sciences, Mumbai 400098, India}

\date{\today}% It is always \today, today,

%%%%%%%%%%%%%%%%%%%%%%%%%%%%%%%%%%%%%%
\begin{abstract}
The origin of the large $\alpha$ particle production and incomplete fusion in reactions involving weakly-bound $\alpha$+$x$ cluster nuclei still remains unresolved. While the (two-step) process of breakup followed by capture of the ``free" complementary fragment ($x$) is widely believed to be responsible,  a few recent studies suggest the dominant role of (direct) cluster stripping.  To achieve an unambiguous experimental discrimination between these two processes, a coincidence measurement between the outgoing $\alpha$ particles and $\gamma$ rays from the heavy residues has been performed for the $^7$Li($\alpha$+triton)+$^{93}$Nb system. Proper choice of kinematical conditions allowed for the first time a significant population of the region accessible only to the direct triton stripping process and not to breakup followed by the capture of the ``free'' triton (from the three-body continuum). This result, also supported by a cluster-transfer calculation, clearly establishes the dominance of the direct cluster-stripping mechanism in the large alpha production. 
\end{abstract}

%\tableofcontents
\end{frontmatter}
%\section{Introduction}
%Introduction: 
Clustering is a general phenomenon observed over a wide range of physical scales and in diverse fields such as the aggregation of galaxies in the universe or the existence of gene clusters in complex biological systems. In nuclear physics, the enormous pairing stability in fermionic quantum systems leads to a large binding energy for the $\alpha$ particle and consequently $\alpha$ clustering is very prevalent in atomic nuclei \cite{Free07}. While the $\alpha$ decay of radioactive nuclei was long ago adduced as direct evidence that $\alpha$ particles formed constituents of heavier nuclei~\cite{Ruth08}, the origin and consequences of $\alpha$ clustering in nuclei remain the subject of intense research due to its importance in fundamental nuclear physics as well as other areas~\cite{AlphaCluster21}. 

In many light nuclei, $\alpha$ clustering is responsible for the weak binding of $\alpha+x$ cluster configurations. A large $\alpha$-particle yield compared to that of the complementary fragment ($x$) is observed in nuclear reactions involving such nuclei, e.g., $^{6,7}$Li and $^{7,9}$Be, indicating that it cannot arise solely due to simple breakup of the weakly-bound projectile in the field of the target nucleus. Capture/transfer of the complementary fragment/cluster, also sometimes referred to as incomplete fusion (ICF), has also been
observed with similar magnitudes in these reactions. Further, the measured large $\alpha$ production and ICF cross sections are commensurate with an observed suppression of the complete fusion (CF) process, suggesting a common origin. However, it is still debated whether the former process influences the multi-dimensional quantum tunneling of fusion in a coherent or incoherent way~\cite{Dasg07,Trip02}. The mechanisms responsible for the large $\alpha$-particle production, ICF and fusion suppression remain unclear and the subject of current interest~\cite{Cook19,Lei19a,Lei19,Rang20,Shri13,Jha20,Bott15,Hodg03}. Unraveling the reaction mechanisms in systems involving nuclei with clustering and weak binding, common features of many light radioactive ion beams (RIBs), is important not only from its fundamental aspect, but also as a promising tool in other areas including nuclear astrophysics \cite{NuAs21,Bert16,Baur04} and nuclear energy applications~\cite{Wang19,Esch12}.
  
Incomplete fusion can arise due to either direct stripping of a cluster from a bound state of the projectile or fusion of one of the ``free'' fragments after breakup of the projectile, i.e.\ so-called breakup-fusion~\cite{Shri13,Trip05,Utsu83,Cast78,Queb71}. Experimentally, it is challenging to distinguish between these two mechanisms as both lead to the same final products with similar energy and angular distributions. Although a clear experimental identification of the underlying mechanism
could not be achieved in earlier coincidence measurements, e.g.~\cite{Shri13,Trip05,Utsu83,Cast78}, a dominant role of breakup-fusion
was suggested by comparing the results with calculations based on a semi-classical model~\cite{Shri13,Torr07} or otherwise~\cite{Trip05}. In a recent study, a few exclusive events ($<$1\% of the total $\alpha$ yields) could be identified as arising from direct stripping only and comparing the inclusive energy-angle distribution with simulations, it was concluded that direct stripping plays a dominant role in ICF~\cite{Cook19}. Thus, a model independent experimental demonstration of the ICF mechanism is still missing. 

\begin{figure}[t]
\includegraphics[trim = 0mm 0mm 0mm 0mm, clip,width=76.0mm]{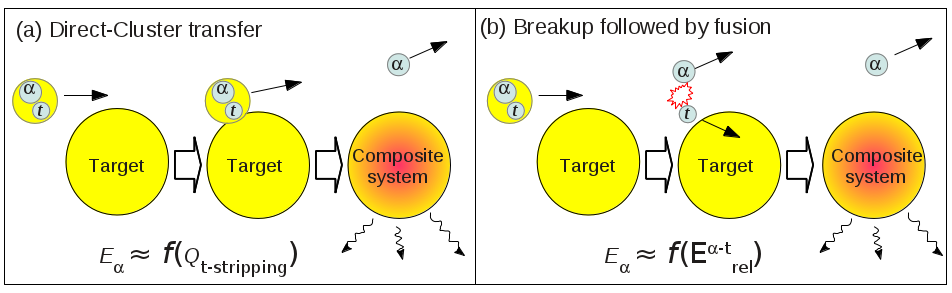} %height=14 cm
\caption{\label{BUvsTransfer} Illustration of (a) direct cluster transfer and (b) breakup followed by fusion of one of the cluster fragments for a $^7$Li($\alpha +t$)+target reaction.}
%\vskip -3mm
\end{figure}

Significant theoretical effort has been invested in understanding the mechanism of the large $\alpha$-particle production and ICF cross sections~\cite{Lei19,Lei19a,Rang20,Park16,Pote15,Torr07,Thom04,Shya91,Aust87}. Recently, using a non elastic breakup model, the cluster-stripping process was shown to be the dominant mechanism for ICF~\cite{Lei19,Lei19a}. In other studies, fusion of the breakup fragments was considered to be the main ICF mechanism~\cite{Rang20,Shri13,Park16,Torr07}. Calculations assuming both cluster stripping~\cite{Lei19,Lei19a} and breakup-fusion~\cite{Rang20,Shri13,Park16,Torr07} mechanisms have successfully reproduced experimental inclusive $\alpha$ yields and/or fusion data to a similar extent. Further, it has also been suggested that  breakup-fusion and transfer to the continuum of the target are equivalent~\cite{Rang20,Pote15}. It is therefore essential to have experimental data populating the bound states and the continuum with comparable magnitude to discriminate between the two widely different mechanisms. As depicted in Fig.\ \ref{BUvsTransfer}, while the outgoing $\alpha$ particle has access to the reaction $Q$ value in the case of direct stripping, the triton fusion (second step) $Q$ value can not be shared with the $\alpha$ particle (produced in the first step) in the two-step breakup-fusion process. A suitable choice of experimental conditions could therefore allow a region which is exclusively populated by only one of these processes to be studied.

This letter reports a measurement of particle-$\gamma$ coincidences for the $^7$Li+$^{93}$Nb system to identify the mechanisms responsible for the origin of the large $\alpha$-particle production and ICF by exploiting the kinematical conditions illustrated in Fig.\ \ref{BUvsTransfer}. Experimental observations are compared with Monte-Carlo simulations and quantum mechanical calculations based on the distorted-wave Born approximation (DWBA) for breakup and cluster transfer, respectively.

%\section{Experimental Details}\label{expt}
%Experimental Details: 

The inclusive and exclusive measurements of $\alpha$ particles were carried out using the $^7$Li beam from the BARC-TIFR Pelletron-Linac facility, Mumbai, in two separate experiments. Targets were self-supporting $^{93}$Nb foils of thickness $\sim 1.6$ mg/cm$^2$. A beam energy of 24 MeV (1.2$V_{\rm B}$) was chosen for both measurements to optimize the kinematical conditions and cross section.
  
In the exclusive measurement, prompt $\gamma$-ray transitions were detected using the Indian National Gamma Array (INGA)~\cite{Pali12}, consisting of 18 Compton suppressed high purity germanium (HPGe) clover detectors. Three Si surface barrier telescopes (thicknesses $\Delta E$ $\sim$ 15-30 $\mu$m, $E$ $\sim$ 300-5000 $\mu$m) were kept at $35^\circ$, $45^\circ$ and $70^\circ$ for the detection of $\alpha$ particles around the grazing angle. One Si surface barrier detector (thickness $\sim 300$ $\mu$m) was fixed at 20$^\circ$ to monitor Rutherford scattering for absolute normalization purposes. The time stamped data were collected using a digital data acquisition system with a sampling rate of 100 MHz \cite{Pali12}. Efficiency and energy calibration of the clover detectors were carried out using standard calibrated $^{152}$Eu and $^{133}$Ba $\gamma$-ray sources. 

In the inclusive measurement, angular distributions of $\alpha$ particles and elastically scattered $^7$Li were measured with three Si surface-barrier detector telescopes (thicknesses: $\Delta E$ $\sim$ 20-50 $\mu$m, $E$ $\sim$ 450-1000 $\mu$m)   mounted on a movable arm inside the scattering chamber. 

\begin{figure}[t]
\centering
\includegraphics[trim = 0mm 2mm 0mm 0mm, clip,width=80.0mm]{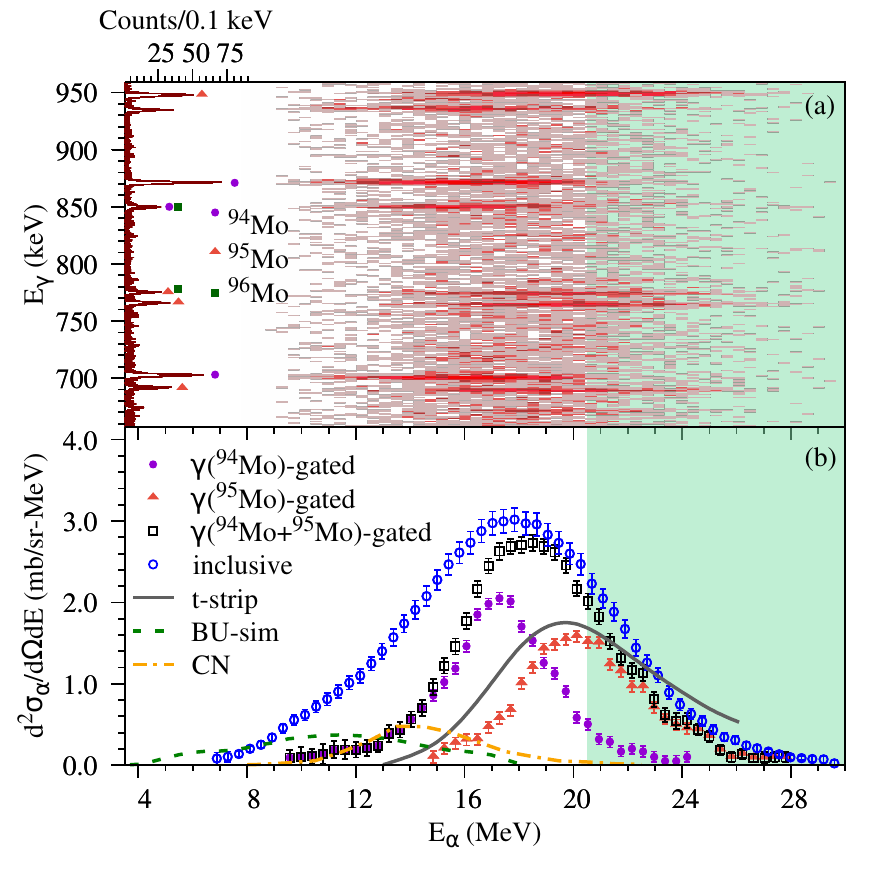}
\includegraphics[trim = 9.4mm 0mm 0.mm 33.mm, clip,width=79.0mm]{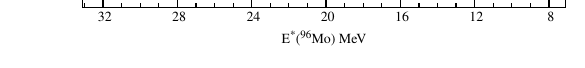}
\caption{\label{EalphaVsEgamma} Particle-$\gamma$ coincidence data: (a) Energy correlation spectrum of prompt $\gamma$ rays and outgoing $\alpha$ particles, detected at an angle $\theta_{\rm lab}$=35$^\circ$ in the $^7$Li+$^{93}$Nb reaction at $E_{\rm beam}$ = 24 MeV. The projection of the energy of the $\gamma$ rays is shown on the left side. The characteristic $\gamma$ rays from residues ($^{94-96}$Mo) formed by $t$ capture are marked. (b) The measured energy spectra of $\alpha$ particles in coincidence with characteristic $\gamma$ transitions from $^{94}$Mo and $^{95}$Mo are shown by circles and triangles, respectively and their sum by open squares. The shaded energy zone ($E_{\alpha}>$20.5 MeV, $E^*<$16.5 MeV) corresponds to stripping of a triton from a bound state in $^7$Li to a bound state in $^{96}$Mo, which is not accessible to breakup followed by ``free'' triton capture. The simulated energy spectrum of $\alpha$ particles due to breakup of $^7$Li involving 3-body kinematics is shown by the dashed line. The dot-dashed line represents the estimated energy spectrum of $\alpha$ particles evaporated from the compound nucleus $^{100}$Ru.}
\end{figure}

%\section{Data Analysis and Results}\label{result}
%Analysis and Results: 
A typical energy correlation spectrum of prompt $\gamma$ rays versus outgoing $\alpha$ particles, detected at an angle of 35$^\circ$, is shown in Fig.~\ref{EalphaVsEgamma}(a). Photo-peaks corresponding to the residues ($^{94-96}$Mo) formed after triton capture ($t$+$^{93}$Nb $\rightarrow$ $^{96}$Mo) are identified and labeled on the projected spectrum in the same figure. Other possible sources of $^{94-96}$Mo residues are compound nuclear evaporation ($\alpha xn$), $d$-stripping: $^{93}$Nb($^7$Li,$^5$He)$^{95}$Mo, and $p$-stripping: $^{93}$Nb($^7$Li,$^6$He$^*$)$^{94}$Mo reaction channels. In case of $p$-stripping, the ejectile ($^6$He) has to be left in an excited state above its $2n$ emission threshold (975 keV) in order to give an $\alpha$ particle in coincidence with a characteristic $\gamma$-rays of $^{94}$Mo. As can be seen, $\gamma$ transitions from $^{96}$Mo are mixed with intense transitions from $^{94}$Mo and $^{95}$Mo. However, after the selection of high energy $\alpha$ groups ($E_{\alpha} >$27 MeV, $E^*<$10 MeV), the dominant yrast transitions of $^{94}$Mo disappear and the weakly populated  photo-peak at 850 keV could be attributed to the $^{96}$Mo (4$^+\rightarrow$ 2$^+$) transition only. This indicates the formation of $^{96}$Mo with excitation energy $E^*<$10 MeV. Production of $^{96}$Mo at such a low $E^*$ via the two-step breakup-fusion mechanism is very unlikely from $Q$-value arguments. Considering the minimum $E^*$=16.5 MeV, corresponding to the $Q$ value for $t$-fusion with $^{93}$Nb, the statistical model code {\footnotesize PACE} predicts $\sim$0.01\% survival probability of $^{96}$Mo against neutron evaporation~\cite{Gavr80}. Hence, the formation of $^{96}$Mo indicates $t$-cluster stripping from a bound state of $^7$Li to a bound state of $^{96}$Mo. A similar observation has been reported for the formation of $^{212}$Po ($t$+$^{209}$Bi) in Ref.~\cite{Cook19}. 

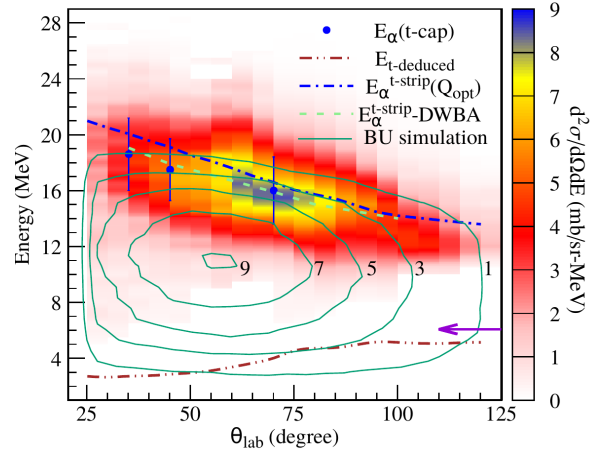
\begin{figure}[t]
\centering
\includegraphics[trim = 2.8mm 0mm 0mm 0mm, clip,width=72.0mm]{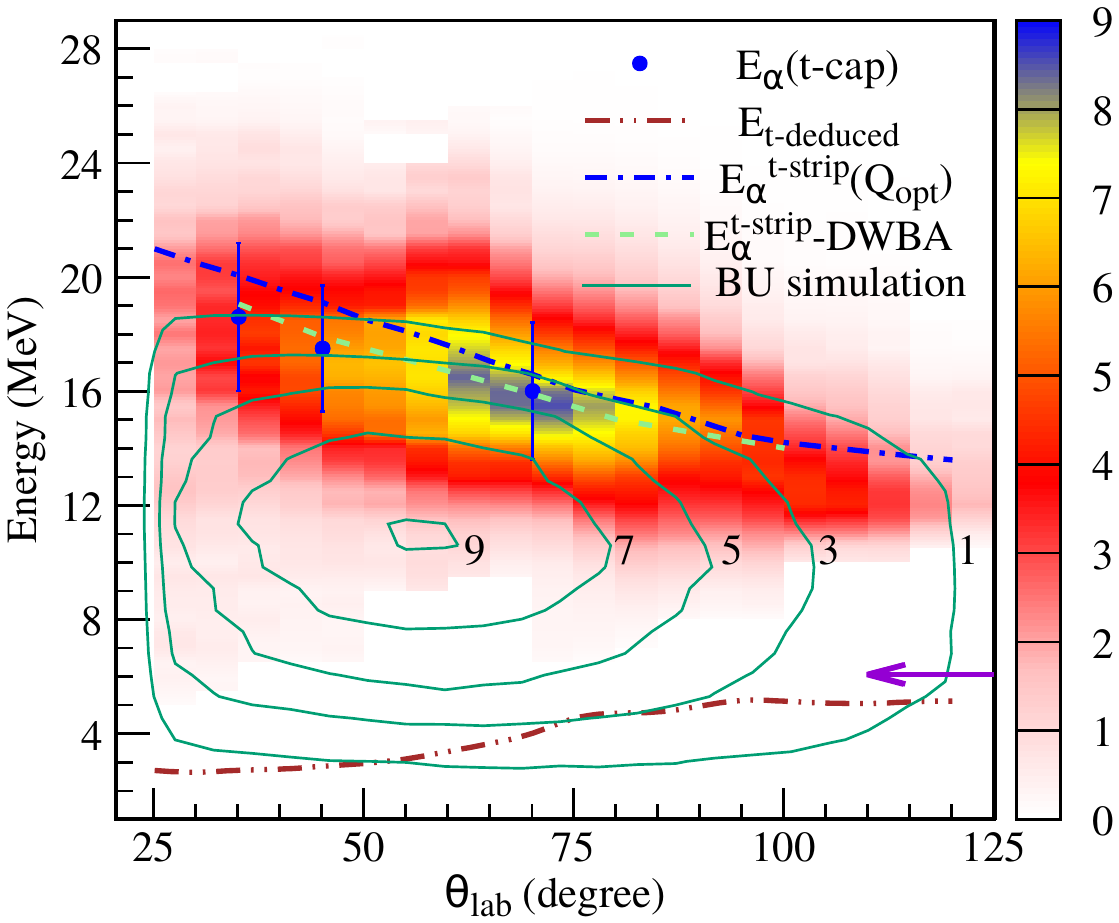}
\begin{picture}(0,00)
\put(-1,130){\rotatebox{-90}{\footnotesize{d$^2\sigma$/d$\Omega$dE (mb/sr-MeV)}}}
\end{picture}
\caption{\label{inclusive} Measured energy-angle correlation spectrum of inclusive $\alpha$ particles in the $^7$Li+$^{93}$Nb reaction compared with the kinematical line for $Q_{\rm opt}$ (dot-dashed line) and the DWBA calculations (dashed line) for $t$ stripping. Mean values of the measured $\alpha$ energy at three different laboratory angles in coincidence with prompt $\gamma$ rays are shown as filled circles along with the width (FWHM). The results of a breakup simulation are shown as contours (see text for details). The energy of the triton calculated from the mean energy of the inclusive $\alpha$ spectra assuming breakup followed by capture of a triton by the target is shown by the dot-dot-dashed line. The arrow indicates the position of the Coulomb barrier between the triton and $^{93}$Nb.}
\end{figure} 

Measured $\alpha$-energy spectra at $\theta_{\rm lab}$=35$^\circ$ obtained by placing $E_\gamma$ gates on the $^{94,95}$Mo residues are presented in Fig.\ \ref{EalphaVsEgamma}(b). The corresponding excitation energy in $^{96}$Mo (assuming a triton capture) is also shown. The sum of the $^{95}$Mo and $^{94}$Mo cross sections is denoted by the open squares and found to explain $\sim$70$\%$ of the measured inclusive yield (shown by open circles). This result is consistent with our earlier conclusions obtained from the $t$-capture residue measurements reported in Ref.\ \cite{Pand17}. The remaining inclusive yield in the low energy range (8-18 MeV) could be  due to breakup of $^7$Li $\rightarrow\alpha+t$ via resonant and non-resonant states, 1$n$-stripping: $^{93}$Nb($^7$Li,$^6$Li$^*$ $\rightarrow \alpha +d$)$^{94}$Nb, 2$n$-stripping: $^{93}$Nb($^7$Li,$^5$Li $\rightarrow \alpha +p$)$^{95}$Nb, and 1$p$-pickup: $^{93}$Nb($^7$Li,$^8$Be $\rightarrow \alpha +\alpha$)$^{92}$Zr. The energies of $\alpha$ particles arising from these reactions, estimated using 3-body kinematics and $Q$-value considerations, are in the ranges of 5-17 MeV ($\alpha + t$), 9-17 MeV ($\alpha + d$), 11-23 MeV ($\alpha + p$) and 13-18 MeV ($\alpha + \alpha$), having a good overlap with the remaining inclusive yield. 

Spectra of $\alpha$ particles from breakup and compound nuclear evaporation were estimated and plotted in Fig.\ \ref{EalphaVsEgamma}(b). The compound nuclear contribution was estimated using the statistical model code {\footnotesize PACE}. The $\alpha$ spectrum due to breakup of $^7$Li was simulated using a {\footnotesize MONTE CARLO} code involving 3-body kinematics, which calculates the energy of the outgoing $\alpha$ particle and triton after breakup. This code can be used to describe the energy-angle characteristics of both breakup fragments, detected simultaneously or in singles mode~\cite{Pand19jinst}. In this simulation, in order to estimate the contribution of breakup to the $\alpha$ + $^{94-96}$Mo  coincidence spectra, it is assumed that after the initial breakup of the $^7$Li ($^7$Li+$^{93}$Nb $\rightarrow \alpha+t +^{93}$Nb) the triton fragment is captured by the target, i.e.\ a breakup-fusion reaction occurs. In such a process the energy of the $\alpha$ particle is independent of the second step, fusion of the triton. The simulated $\alpha$-particle energy spectrum from breakup is thus the mixture of non-capture breakup and breakup-fusion events. In the simulation, relative energies between the $\alpha$ and triton up to 10 MeV were considered and fragment emission in the rest frame of the $^7$Li was assumed to be isotropic. The angular distribution of the scattered $^7$Li$^*$($\rightarrow \alpha +t$) was estimated by a continuum-discretized coupled-channels (CDCC) calculation, using the code {\footnotesize FRESCO}~\cite{Thom88}. The $\alpha + t$ continuum of $^7$Li was divided into bins in momentum ($k$) space of width $\Delta k = 0.1$ fm$^{-1}$ up to $k_\mathrm{max} = 0.8$ fm$^{-1}$. Bins with relative angular momenta between the clusters of $L=0$--$4$, together with the $7/2^-$ (4.63 MeV) and $5/2^-$ (6.68 MeV) $L=3$ resonance states, were included in the calculations with couplings (including continuum-continuum couplings) up to multipolarity $\lambda = 4$. The $\alpha$+$^{93}$Nb and $t$+$^{93}$Nb optical potentials required as input to the Watanabe-type folding potentials were taken from Refs.\ \cite{Avri03} and \cite{Cho07}, respectively. The real and imaginary parts of the resulting folded potentials were normalized by factors of 0.75 and 0.70 respectively to give a good description of the elastic scattering data by the full CDCC calculation.

As can be seen, the characteristics of the simulated breakup spectrum are very different from the exclusive data. In particular, the shaded region $E_{\alpha}>$20.5 MeV ($E^*<$16.5 MeV) can not be populated by the breakup process. Moreover, this zone also corresponds to bound $t$+$^{93}$Nb configurations in $^{96}$Mo. Hence, the measured coincident $\alpha$ yields above $E_{\alpha}=$20.5 MeV must be exclusively due to direct $t$-cluster stripping from bound states of $^7$Li to bound states (with respect to triton emission) of $^{96}$Mo, allowing us to put a lower limit of $\sim$30\% on this contribution. The spectrum below $E_{\alpha}= $ 20.5 MeV could have a component due to breakup mixed with stripping to resonant $t$+$^{93}$Nb states of $^{96}$Mo that decay by neutron emission. However, the shape of the measured spectrum does not support the possibility of a significant contribution from breakup.

To investigate further, the energy-angle correlations of the $\alpha$ particles were analyzed. The experimental correlations obtained from the inclusive and exclusive measurements are compared with those estimated as due to direct stripping and breakup in Fig.~\ref{inclusive}. The measured inclusive and exclusive correlations exhibit similar characteristics. While the experimental correlations are in good agreement with both the kinematical curve using the optimum $Q$ value~\cite{Schi73} and DWBA calculations (discussed later) for $t$-stripping over a wide angular range (25$^\circ$-125$^\circ$), the result of the breakup simulation has very different characteristics. This further demonstrates that breakup does not make a significant contribution to the $\alpha$-particle production and ICF. The energy of the triton ($E_t$) deduced from the measured mean values of $E_\alpha$ assuming the $\alpha$-$t$ breakup mechanism is found to be less than the Coulomb barrier between the $t$-fragment and $^{93}$Nb (indicated by the arrow) over the whole angular range. This also indicates that, due to the fusion barrier, breakup followed by fusion of the triton is very unlikely. This systematic investigation establishes the dominance of cluster stripping over breakup-fusion as the main source for the large $\alpha$ yields and ICF.

\begin{figure}[t]
\includegraphics[trim = 0.5mm 2mm 2mm 9mm, clip,width=78.0mm]{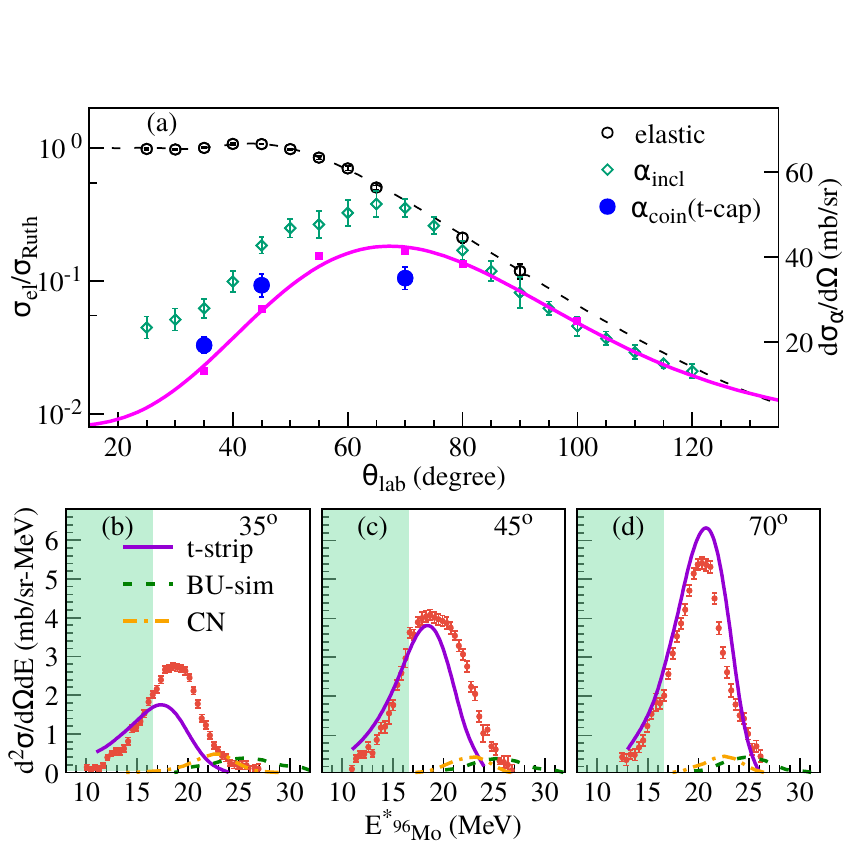} %height=14 cm
\caption{\label{elastic} (a)  Measured elastic scattering and $t$-capture angular distributions compared with the DWBA calculations. The solid line is a smooth fit to the calculated cross sections (filled squares) at discrete angles. Double differential cross sections at (b) 35$^\circ$, (c) 45$^\circ$ and (d) 70$^\circ$ are also compared with the DWBA calculations, compound nuclear contribution and breakup simulation.}
\end{figure}

%\section{Calculation}\label{cal}
%Calculation: 
Since it is evident from the above experimental observation that cluster stripping dominates over breakup-fusion, one-step $t$-stripping calculations were carried out within the DWBA framework using the code {\footnotesize FRESCO}~\cite{Thom88}. The triton was assumed to be transferred as a single entity from the $^7$Li ground state to a $t$ + $^{93}$Nb state of $^{96}$Mo. States above the $t$+$^{93}$Nb separation energy of $^{96}$Mo were treated with a weakly bound approximation method. The calculated cross sections for several small binding energies were extrapolated to the actual positive ``binding energy''. Levels at excitation energies below the triton emission threshold were treated in the usual way for bound states. The entrance channel optical potentials were obtained by fitting the measured elastic scattering, starting from global $^7$Li parameters~\cite{Cook82}. Only minor adjustments to $V$, $W$ and $r_W$ were required, yielding potential parameters: $V = 106.1$ MeV, $r_V = 1.286$ fm, $a_V = 0.853$ fm, $W = 13.36$ MeV, $r_W = 1.739$ fm and $a_W = 0.759$ fm. The exit channel $\alpha$ particle optical potential used a regional parameter set for A $\sim$ 100 nuclei \cite{Avri03}. The measured $j$-distribution from the side feeding in the $\gamma$-ray cascade of $^{94}$Mo, which contributes more than 50$\%$ of the $t$-capture cross section, is found to be peaked at $j$=6. Hence, orbital angular momenta associated with the triton-$^{93}$Nb motion up to $\Delta$L=8 were considered. Spectroscopic factors for $\left<^7\mathrm{Li}|\alpha+t\right>$ and $\left<^{96}\mathrm{Mo}|^{93}\mathrm{Nb}+t\right>$ overlaps were taken as unity. The calculations were normalized to give the best fit to the energy integrated angular distribution. 

Elastic scattering and $t$-stripping angular distributions are compared with the data in Fig.~\ref{elastic}(a). Calculated double differential cross sections are compared with the data in Fig.~\ref{elastic}(b-d). The measured angular distribution and excitation energy spectra are fairly well explained. The observed differences could be due to the limitations of the weakly bound approximation approach and/or contributions from other processes. Since not only the magnitudes (much smaller) but also the peak positions of the simulated breakup spectra are very different from the measured ones, any substantial  contribution from breakup can be precluded. Compound nuclear contributions are shown in Fig.~\ref{elastic}(b-d) and also found to be very small. A significant contribution from $d$-stripping followed by decay $^5$He ejectile may be ruled out on kinematic grounds. The optimum $Q$ value for this reaction is $Q_\mathrm{opt} \approx-7$ MeV and the band of kinematically allowed values of E$_\alpha$ resulting from decay of the $^5$He ground state falls below the measured inclusive $\alpha$-particle energy-angle correlation spectrum in Fig.\ \ref{inclusive} for $Q = Q_\mathrm{opt}$. As discussed in Ref.\ \cite{Pand17}, a theoretical estimate of the $p$-stripping cross section populating the $2^+$ (1.8 MeV) resonance state in  $^6$He is at most a few mb (measured $\sigma(^6$He$_{\rm g.s.}$) $\sim$7 mb \cite{Pand19}), of the order of 2\% of the total inclusive $\alpha$ yield, so that a significant contribution from this process is also ruled out. Simulated $\alpha$ particles spectra produced due to breakup of of the $^6$He ($2^+$) for $Q = Q_\mathrm{opt}$ ($Q_\mathrm{opt} \approx -7$ MeV) also excludes the possibility of a significant contribution, in the same way as for the $d$-stripping.

The present work has clearly identified using kinematical conditions a region of the double differential cross section populated exclusively by direct triton-stripping from the bound states of $^7$Li to bound state of $t$+$^{93}$Nb in $^{96}$Mo which predominantly decay by neutron and/or $\gamma$ emission, denoted by the shaded region in Fig.~\ref{EalphaVsEgamma}(b) and Fig.~\ref{elastic}(b-d). This is direct experimental evidence in support of the results from the theoretical framework of Ref.~\cite{Lei19,Lei19a}, where it is shown that direct capture/stripping from the projectile ground state is the dominant mechanism.
Furthermore, the remaining unshaded region is also mainly due to the direct stripping of triton leading to  formation of $^{96}$Mo$^*$ composite system (with excitation energy above triton separation energy), as the contribution from the breakup of $^7$Li followed by fusion of triton and compound nuclear processes are found to be negligible.

Although, the conclusion of the present investigation is specific to $^7$Li, it can be generalised to other weakly bound stable and radioactive nuclei with predominant cluster structure.

%\section{Summary and Conclusions}\label{sum}
%Summary and Conclusions: 
 In summary, the origin of the large $\alpha$ production and ICF in reactions with a weakly-bound projectile has been investigated by employing the particle-$\gamma$ coincidence method in the $^7$Li+$^{93}$Nb system at a near-barrier energy. With proper choice of kinematical conditions it has been possible for the first time to populate with significant strength the part of the spectrum accessible to triton-cluster stripping only and not to breakup. This provides direct experimental evidence for the dominant role of the stripping process and also allows a meaningful comparison with theoretical models. Double differential cross sections are reproduced well by transfer (DWBA)) calculations. The difference between the observed and simulated breakup spectra rules out any significant contribution from the breakup mechanism to the large $\alpha$ production or ICF, necessitating a revisitation of the role of the continuum in reactions involving weakly-bound projectiles. This unique dataset of cluster-stripping to bound and unbound states of comparable magnitude offers a test-bench for further development of state of the art theoretical formalisms (e.g.~\cite{Lei19,Rang20,Torr07}) for reactions involving weakly-bound stable/radioactive nuclei, which are also used to simulate nucleosynthesis.

We acknowledge the accelerator staff for the smooth running of the machine. Help during the experiment from P. Patale is also acknowledged. One of the authors (S.K.P.) thanks Prof. V. Jha for fruitful discussions.
%\bibliography{/home/sanat/bibFile/7li2021jan}% Produces the 

\end{document}